\documentclass[prb,twoside,showpacs,twocolumn]{revtex4}

\usepackage{graphicx}
\usepackage{color}
\usepackage{epsfig}
\usepackage{amsmath}
\usepackage{amstext}
\usepackage{amssymb}
\usepackage[colorlinks=true,linkcolor=blue]{hyperref}

\begin{document}

\title{Quantum holonomies with Josephson-junction devices}
\author{Mateusz Cholascinski}
\affiliation{Institut f\"ur Theoretische Festk\"orperphysik,
Universit\"at Karlsruhe, D-76128 Karlsruhe, Germany}
\affiliation{Nonlinear Optics Division, Adam Mickiewicz University, 61614 Poznan, Poland}

\begin{abstract}
We examined properties of a Josephson-junction system composed of
two coupled Cooper-pair boxes (charge qubits) as a candidate for
observation of quantum holonomies. We construct a universal set of
transformations in a twofold degenerate ground state, and discuss
the effects of noise in the system.
\end{abstract}


\maketitle


\section{Introduction}\label{sec:intro}

Appearance of the geometric phases in physical systems has been,
since their first systematic treatment by \textcite{berry}, among the
most fascinating physical phenomena. Generalized to systems with
degenerate spectrum,\cite{WZ} they give the possibility to devise systems
in which the dynamical contribution is only the overall phase factor, and the
actual transformations are of purely geometric origin. \\
Quantum geometric transformations (below referred to as {\em
holonomies}) have attracted even more attention after
\textcite{pachos} proved their potential use in quantum computing. As
compared to the ordinary,  
dynamical computation, quantum gates are here realized by cyclic
evolution of parameters, and the result depends only on the geometry of
the traversed path.

Construction of holonomies has been discussed for various physical  
realizations, among them are also superconducting nanocircuits.
\cite{faoro,choi} In these realizations the necessary number of
independent tunable parameters results in high complexity of
the considered systems. Here, motivated also by the recent experiment
with Josephson-junction system composed of two coupled charge qubits,
\cite{pashkin} we consider a similar design as a
potential candidate for 
observation of quantum holonomies. As compared to the proposals in
Refs.~\cite{faoro,choi}, where the simplest two-dimensional holonomies
are constructed using four coupled charge qubits, we achieve substantial
simplification using only two qubits. Also, we realize the transformations
within a twofold  degenerate ground state, rather than excited state (as in
\cite{faoro,choi}). In this way we avoid the problem of depopulation
of the subspace in which the holonomies are realized.

We begin with describing the system and defining the operational
subspace. In the four-dimensional space of two charge qubits we find a
configuration of parameters for which the ground state is
degenerate. We perform the holonomies by selecting two of the
parameters and varying them adiabatically along a certain
loop. During the transformation the degeneracy of the ground state is
maintained by adjusting the remaining parameters as a function
of the others.

The scheme presented here may be realized in a system
without strong constraints on its parameters provided that the noise
level is low. However, since the system of charge qubits is usually
affected by the 
charge fluctuations, in the following we optimize the design in order
to suppress the noise (for certain models of errors
we find a decoherence-free subspace \cite{duan,zanardi,lidar}). 
Finally we discuss possible extensions of the scheme. The specific
system considered here is described by the Hamiltonian of two coupled
qubits. In the discussion we also comment on the possible application of our
scheme to an arbitrary system with this model Hamiltonian.


\section{The system}\label{sec:system}

\begin{figure}[h] 
  \centerline{\resizebox{0.3\textheight}{!}{\rotatebox{0}
      {\includegraphics{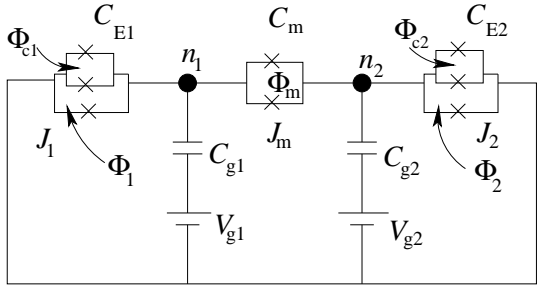}}}}
  \caption{The system of two coupled charge qubits used to construct
      the simplest non-Abelian holonomies. The superconducting islands
      ("Cooper-pair boxes") are denoted by black nodes. $n_1$ and
      $n_2$ are the numbers of excess Cooper pairs on the islands. The
      Josephson coupling 
      between the two islands may be turned off completely. The couplings
      $J_1$ and $J_2$ can be on one hand turned off completely, on the
      other hand we may obtain complex couplings by changing the
      fluxes $\Phi _{c1(2)}$.}
  \label{fi:system}
\end{figure}

The system we consider consists of two
``charge qubits''.\cite{makhlin} They are coupled to each other via
tunable Josephson junction -- a symmetric dc-SQUID (superconducting
quantum interference device) (see
Fig.~\ref{fi:system}). As discussed later, we need to have the
possibility to switch off all the Josephson energies completely (and we
use dc-SQUIDS instead of simple junctions). On the
other hand, to construct nontrivial transformations, the amplitude
and phase (or 
equivalently the real and imaginary parts) of the couplings
$J_1$ and $J_2$ should be controllable during the
operations. To achieve this we replace one of
the junctions in the left and right SQUIDs with a further
dc-SQUIDs (the same technique has been used in Ref.~\cite{faoro}). 

If the dimensionless gate charges $n_{g1(2)} = C_{g1(2)}V_{g1(2)}/2e$
are close to $1/2$, the only relevant charge states for each island are $|0
\rangle $ and $|1 \rangle $ (zero or one extra Cooper pair on the
island). The Hamiltonian may be then reduced to the four-dimensional
space and in the charge basis $|n_1, n_2 \rangle $ has the form
\begin{equation}
  H = \left(
    \begin{array}{cccc}
      E_{00} & -J_2/2  & -J_1/2  & 0  \\
      - J_2^*/2 & E_{01} & -J_m/2 & -J_1/2 \\
      - J_1^*/2 & -J_m/2  & E_{10} & -J_2/2  \\
      0 & -J_1^*/2  & -J_2^*/2 & E_{11} 
    \end{array}
  \right).
  \label{eq:7ra}
\end{equation}

Here the diagonal elements are the electrostatic terms $E_{n1, n2} =
E_{c1} (n_{g1} - n_1)^2 + E_{c2}(n_{g2}-n_2)^2 + E_m
(n_{g1}-n_1)(n_{g2}-n_2)$ [$n_1 (n_2)$ is the number of excess Cooper
pairs on the first (second) island] with $E_{c1} = 4e^2 C_{\Sigma
  2}/2(C_{\Sigma 1}C_{\Sigma 2} - C_m^2)$, and similarly for $E_{c2}$.
Here $C_{\Sigma 1(2)}$ is
the sum of all capacitances connected to the first (second)
island. The electrostatic coupling term due to finite capacitance of
the middle dc-SQUID equals $E_m = 4e^2 C_m/ (C_{\Sigma 1} C_{\Sigma 2}
- C_m^2)$ (we do not assume weak electrostatic coupling and the
capacitance $C_m$ does not need to be small). The Josephson coupling  
$J^{(1)} = \sqrt{(J^{(1)}_{jl}-J^{(1)}_{jr})^2 +
  4J^{(1)}_{jl}J^{(1)}_{jr}\cos^2(\pi \Phi _{1})} \exp [- i \psi
(\Phi _{1})]$ (and similarly for $J^{(2)}$), where $\tan \psi (\Phi_1
) = (J^{(1)}_{jl}-J^{(1)}_{jr})/(J^{(1)}_{jl} + J^{(1)}_{jr}) \tan \pi
\Phi _1$.\cite{Tinkham} Here $J_{jl}$ is the Josephson energy of
the junctions in the dc-SQUIDs. The term 
$J_{jr} = J_{jr (0)} \cos \pi \Phi _{c}$ is the
tunable Josephson coupling of the small SQUID. Finally $J_m =
J_m^{(0)} \cos \pi \Phi _m$ is the Josephson energy of the middle
SQUID. Here the fluxes are in units of $\Phi _0 = h c/ 2 e$, the
superconducting flux quantum. We assume that the system is operated in
the charge regime ($E_{c1(2)}, E_m \gg J_{jl(r)}^{(1)},
J_{jl(r)}^{(2)}, J_m^{(0)}$).

A similar system has been used already in the context of geometric
phases. \textcite{falci} used two coupled charge qubits to devise an
experiment in which the Berry phase could be detected. Also, as
already mentioned, the
experiment by \textcite{pashkin} shows that a system of two coupled
charge qubits can be well controlled and manipulated with available
experimental techniques. The difference between our system and those
used in Refs. \cite{falci,pashkin} is that we couple the islands using
dc-SQUID instead of a capacitor, and we require controllable Josephson
couplings $J_{1(2)}$.

The set of controllable parameters defines the {\em control
  manifold}. The Hamiltonian is characterized by the tunable
terms: $E_{c1}$, $E_{c2}$, $J_1$, $J_2$, $J_m$. We will use in the
discussion rather the parameters that can be controlled directly,
{\it i.e.}, $(\Phi _1, \Phi _2, \Phi _m, n_{g1}, n_{g2})$. The fluxes
$\Phi _{c1(2)}$ are used only to control the
symmetry of the SQUIDs, and we will not consider them to be used as the actual
tunable parameters in the following discussion (between the operations
they are kept at the values for which the couplings $J_{1(2)}$ are
real, and may be tuned to zero. Just before the operations we can
switch them to different values for which the dc-SQUIDs are
asymmetric. During the operations they are kept constant).

The operational subspace for the simplest non-Abelian holonomies is
spanned by two lowest-energy eigenstates. To make the transformations
of purely geometric nature we need to make them degenerate. This
property in our system does not follow from the symmetry arguments
and we need to control it by making the parameters not fully
independent. In the following we will compare the energies of the
lowest states, and from this constraint calculate one of the
parameters as the function of the remaining ones. In this way we
define the degeneracy domain in the control manifold.

For the following configuration of the parameters
\begin{equation}
  J_m = J_1 = J_2 = 0, \quad n_{g1} = n_{g2} = 1/2,
  \label{eq:qra}
\end{equation}
the Hamiltonian Eq.(\ref{eq:7ra}) is diagonal in the charge basis, and
the ground state is twofold degenerate and spanned by the states
\begin{equation}
  |\bar{0} \rangle \equiv |01 \rangle, \quad |\bar{1} \rangle \equiv
  |10 \rangle.
  \label{}
\end{equation}
We will refer to this particular point in the parameter space as the
{\em starting point}, or the ``no-op'' regime (as this configuration
will be kept between the sequences of transformations), and to the
states $|\bar{0} \rangle $ and $|\bar{1}
\rangle $ as the logical basis. 

Before we perform any transformations the system is initialized to the
state $|\bar{0} \rangle $ by turning off all the
Josephson couplings, and tuning the electrostatic energy $E_{01}$ to
be the lowest one. After long enough time the system will relax, and
we switch to the starting point. 
We perform a holonomic transformation by varying the parameters
adiabatically along a suitable loop. 
During the operations the other charge states may be involved as
well. The system,
however, at any instant of time remains in the twofold degenerate
ground state, and
the resulting unitary transformation 
(holonomy) is limited to the logical basis only. The correspondence
between the  loop, and the performed unitary transformation is given by
\begin{equation}
  U_{\Gamma } = {\cal P} \exp
  \left(- \oint_{\Gamma } \sum_{i} {\cal A}_i d X_i \right),
  \label{}
\end{equation}
where ${\cal P}$ is the path-ordering operator, $\Gamma$ the
 path, $R = \{X_i\}$ is the set of parameters (control
manifold), and the $2 \times 2$ 
matrix ${\cal A}_i$ is the {\em Wilczek-Zee connection},\cite{WZ}
\begin{equation}
  {\cal A}_{i}^{\alpha \beta } = \langle \alpha (R) |
  {\partial\over{\partial X_i}} |\beta (R)  \rangle  .
  \label{}
\end{equation}
$|\alpha  \rangle$ and  $|\beta  \rangle  $ are
parameter-dependent states spanning the operational subspace, and for the
'no-op' regime they can be either $|\bar{0} \rangle $ or $|\bar{1}
\rangle $.

To construct the holonomies we need to find a
loop corresponding to the desired transformation. 
Any unitary transformation can be factorized as a product of
two noncommuting rotations in the qubit space. Here we find two
such holonomies, namely, $\exp (i
\phi _1 \sigma _z)$ and $\exp(i \phi _2 \sigma _x)$. For our purposes
it is convenient to interpret these rotations as phase shifts
between $|\bar{0} \rangle $ and $|\bar{1} \rangle $,
and between $|+ \rangle = \left(|\bar{0} \rangle + |\bar{1} \rangle \right) /
\sqrt{2}$ and $|- \rangle = \left(|\bar{0} \rangle - |\bar{1} \rangle \right) /
\sqrt{2}$, respectively. For given rotation, at each point of the
constructed paths  
the Wilczek-Zee connection ${\cal A}$ is then diagonal [in the basis
$|\bar{0} (\bar{1}) \rangle $ and $|+ (-) \rangle $ respectively], and we
omit the path-ordering operation.


\section{Constructing the holonomies}\label{sec:transformations}

\begin{figure}[h] 
\centerline{\resizebox{0.29\textheight}{!}{\rotatebox{0}
{\includegraphics{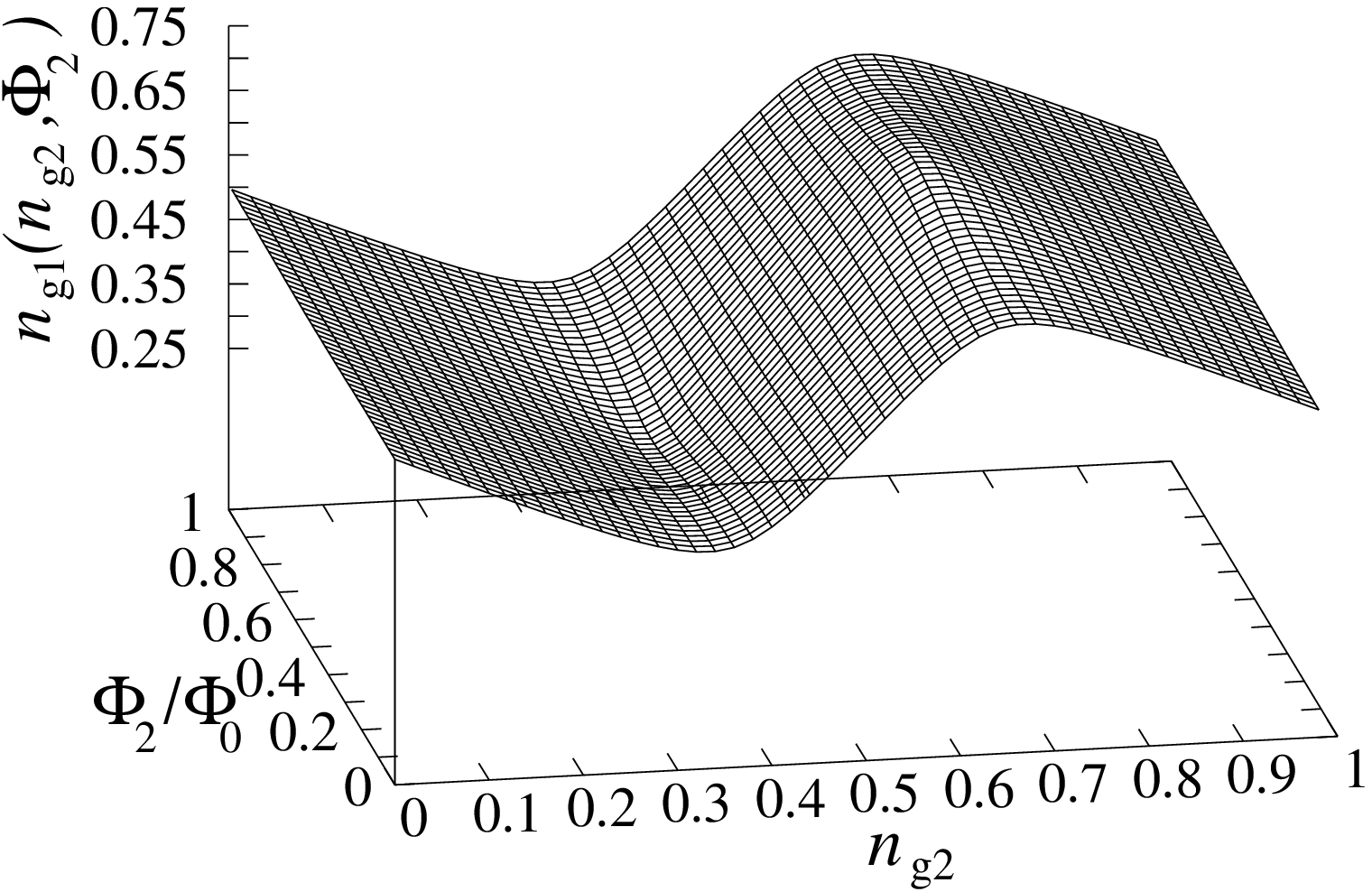}}}}
\centerline{\resizebox{0.29\textheight}{!}{\rotatebox{0}
{\includegraphics{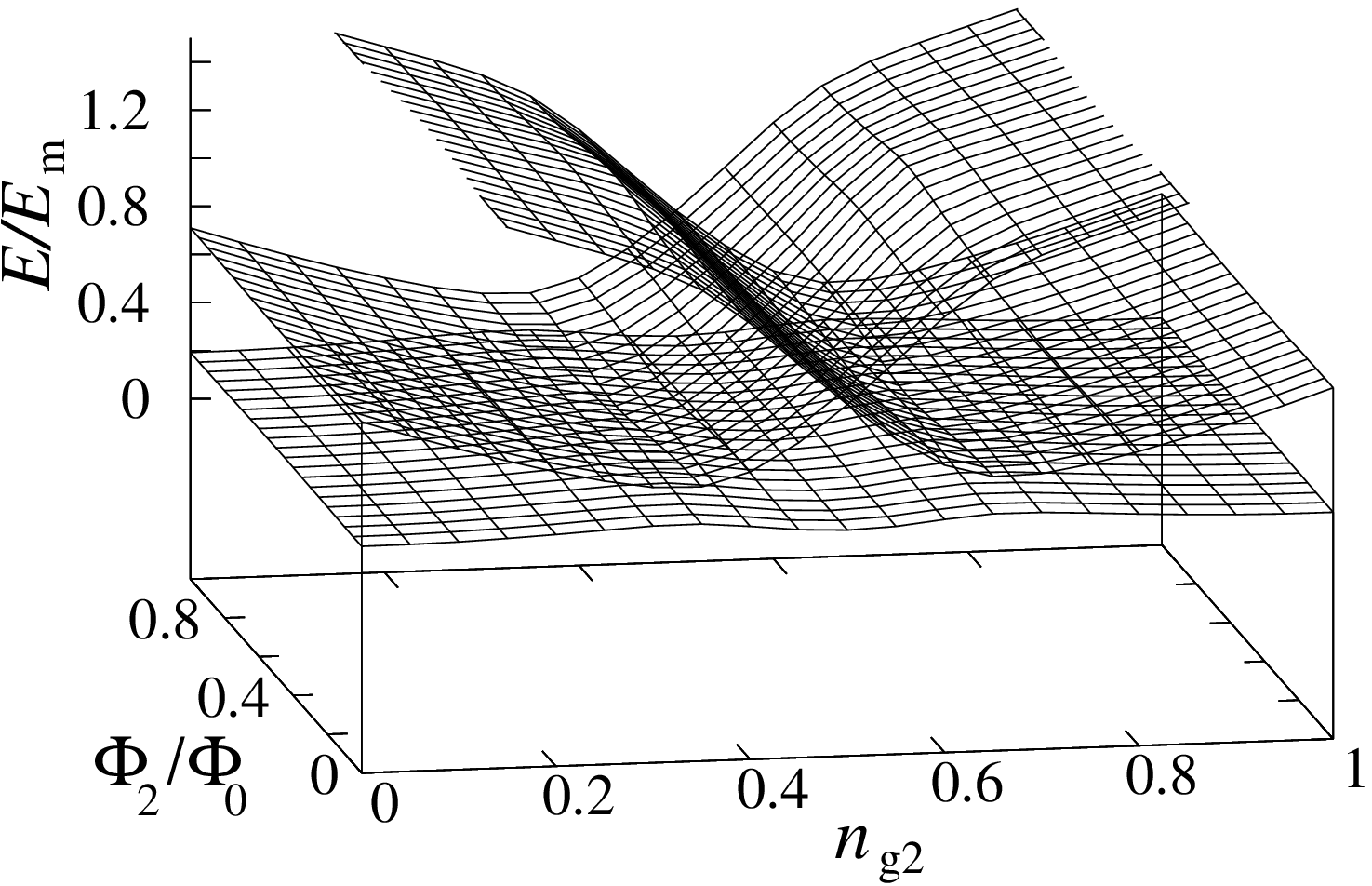}}}}
\centerline{\resizebox{0.29\textheight}{!}{\rotatebox{0}
{\includegraphics{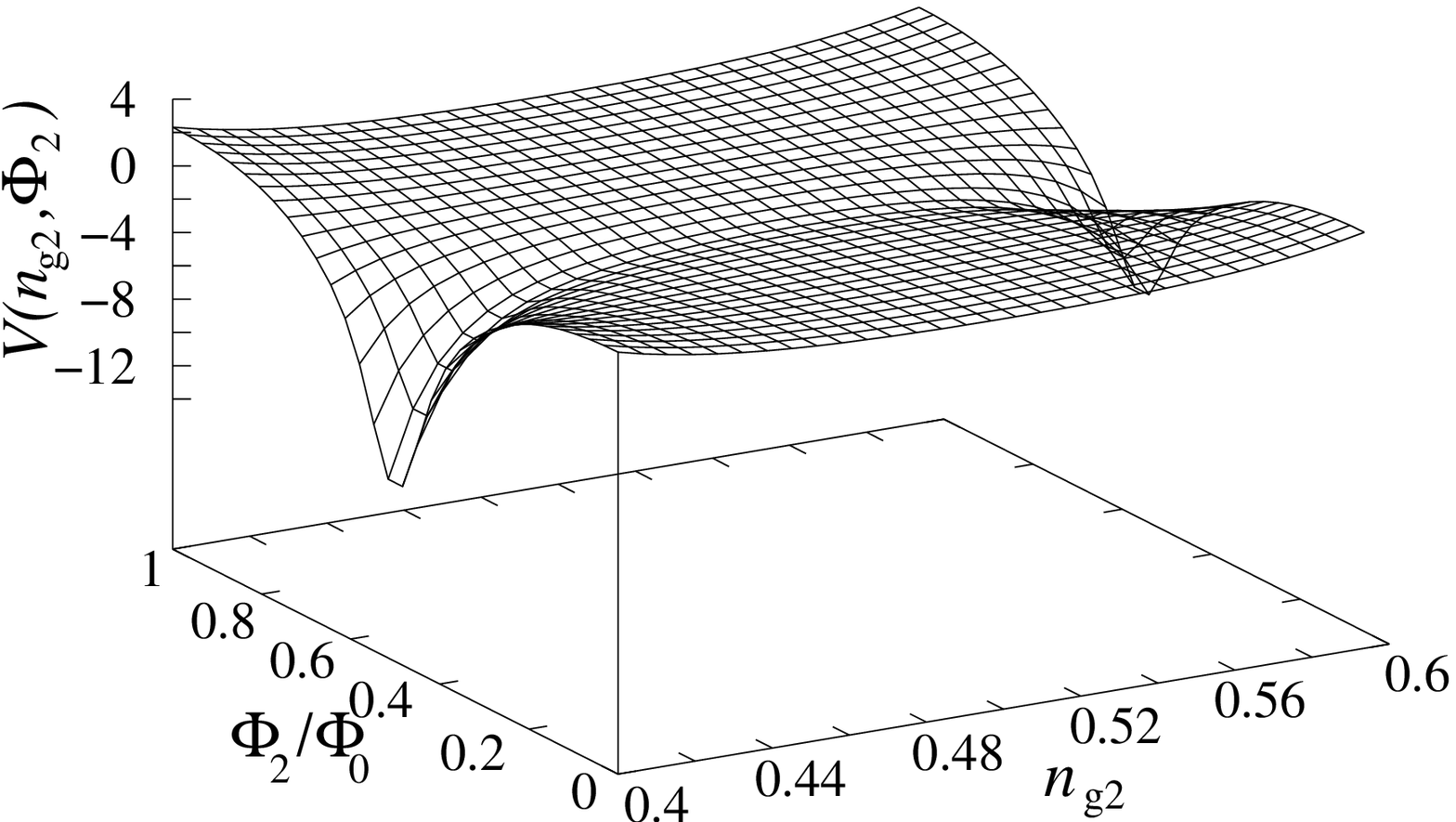}}}}
\caption{Construction of the $\exp (i \phi _1 \sigma _z)$
  rotation. Upper plot -- the degeneracy subspace calculated from the
  comparison of the energies of the lowest-energy levels; middle
  plot -- the parameter-dependent spectrum of the system (the ground
  state is twofold degenerate); lower plot -- the ``Berry field''. All
  plots taken for the following values of system parameters:
  $E_{c1}/E_m = 0.8$, $E_{c2}/E_m = 1$, $J_{r}^2/E_m
  = 0.1$, and $J_{l}^2/E_m = 0.05$.} 
\label{fi:plotsphase}
\end{figure}

The holonomies we find are in our case the aforementioned rotations, $e^{i
\phi _1 \sigma _z}$ and $e^{i \phi _2 \sigma _x}$. 
The first transformation we consider is the phase shift $e^{i \phi _1
  \sigma _z}$. If we keep $J_1 = J_m \equiv 0$ (and change the
flux $\Phi _{c2}$ to make the coupling $J_2$ complex), the Hamiltonian
simplifies to  
\begin{equation}
  H_1 = \left(
    \begin{array}{cccc}
      E_{00} & -J_2/2  & 0  & 0  \\
      - J_2^*/2 & E_{01} & 0 & 0 \\
      0 & 0  & E_{10} & -J_2/2  \\
      0 & 0 & -J_2^*/2 & E_{11}
    \end{array}
  \right).
  \label{}
\end{equation}
The obtained block form may be solved for each part independently. We
calculate the desired phase shift as the difference between the Berry
phases for each block. At the same time, to make the phase shift of
purely geometric origin, we compare the energies of the lower states
within each block to find the degeneracy subspace in the control
manifold. Since the energies of the lowest states are functions of
$n_{g1}, n_{g2}$, and $\Phi _2$, from the condition $E_{\alpha
}(n_{g1}, n_{g2}, \Phi _2) - E_{\beta }(n_{g1}, n_{g2}, \Phi _2) = 0$
we find the degeneracy condition in the functional form $n_{g1}
(n_{g2}, \Phi _2)$. This dependence is shown in the upper part of
Fig.~\ref{fi:plotsphase}

The parameter-dependent spectrum (for the degeneracy domain in the
control manifold) is
shown in the middle plot of 
Fig.~\ref{fi:plotsphase}. The effectively four-level system has only
three distinct energy values due to the degeneracy of the ground
state. 

Finally the phase shift may be presented conveniently as the 
integral,
\begin{equation}
  \phi _1 =  - {1\over 2} \int_{S(\Gamma )} d n_{g2} d \Phi _2
  \left[V_{\alpha } 
  (n_{g2}, \Phi _2) - V_{\beta }
  (n_{g2}, \Phi _2)\right],
  \label{}
\end{equation}
where $S(\Gamma )$ is the domain in the $n_{g2}$-$\Phi _2$ plane
limited by the path $\Gamma $, and 
the ``Berry field'' is given by \cite{berry}
\begin{equation}
  V_{\alpha (\beta )} = - i {\langle \alpha (\beta ) | \nabla_{R} H_1
    |e_{a} (e_b) \rangle \times \langle e_{a} (e_b) | \nabla_{R} H_1
    |\alpha  (\beta ) \rangle \over (E_{\alpha (\beta ) } - E_{ea (eb)})^2}.
  \label{}
\end{equation}
Here $|e_a \rangle $ and $|e_b \rangle $ are the excited states
from the upper-left and the lower-right block of $H_1$ respectively, and $E_{ea
  (eb)}$ their energies. The field difference $V = V_{\alpha } -
V_{\beta }$ is shown in the lower plot of Fig.~\ref{fi:plotsphase}.

\begin{figure}[h] 
\centerline{\resizebox{0.29\textheight}{!}{\rotatebox{0}
{\includegraphics{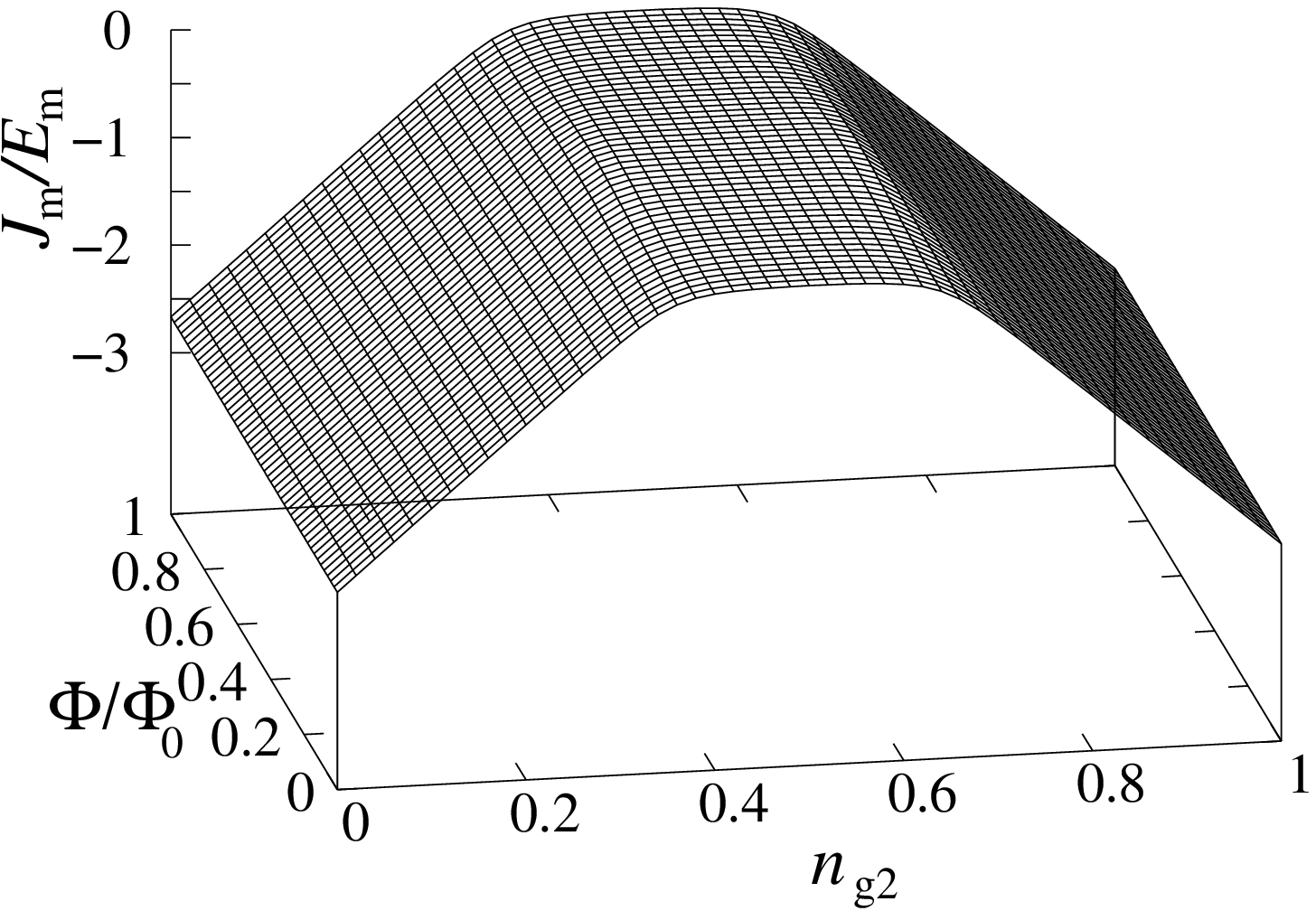}}}}
\centerline{\resizebox{0.29\textheight}{!}{\rotatebox{0}
{\includegraphics{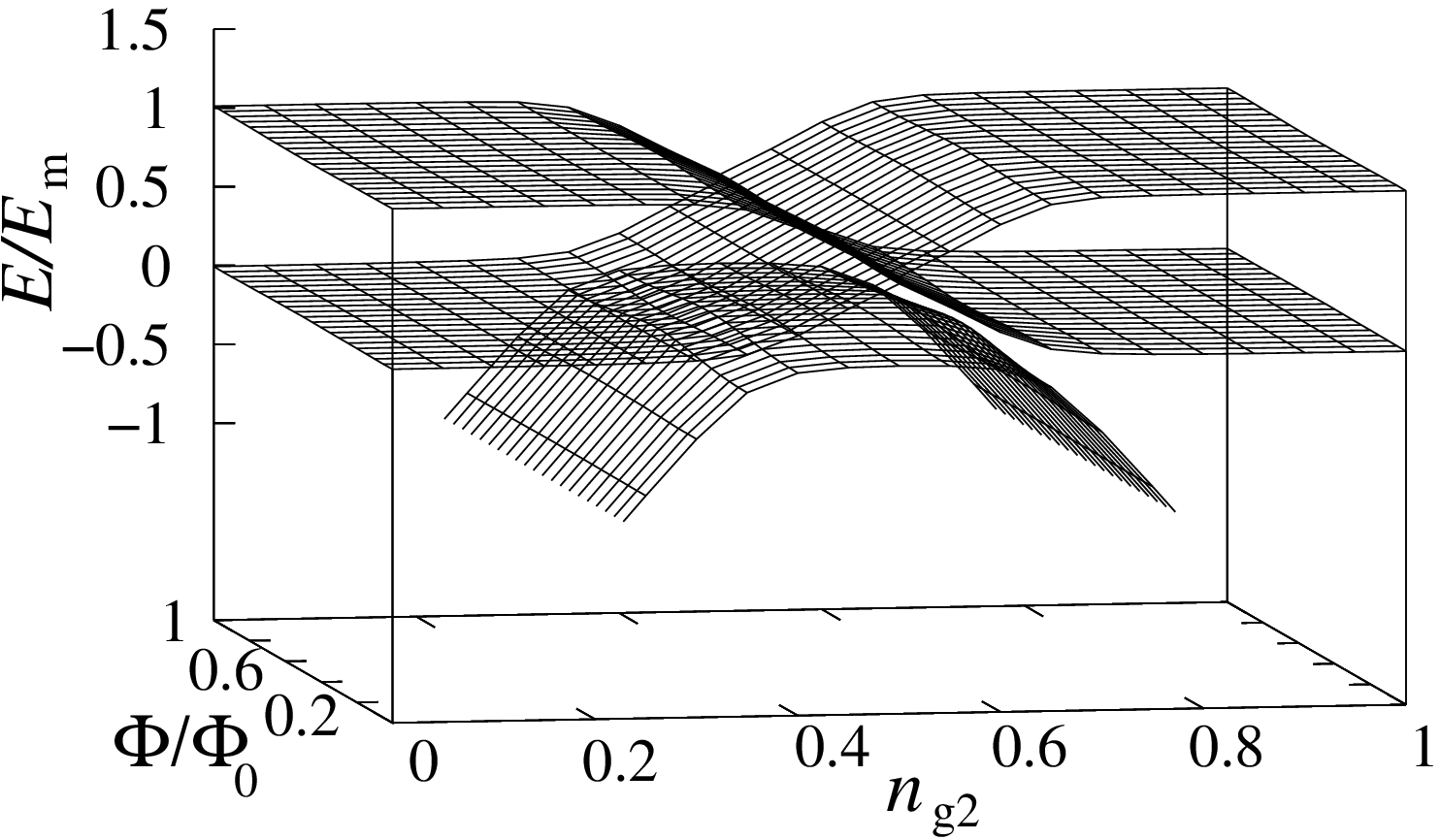}}}}
\centerline{\resizebox{0.29\textheight}{!}{\rotatebox{0}
{\includegraphics{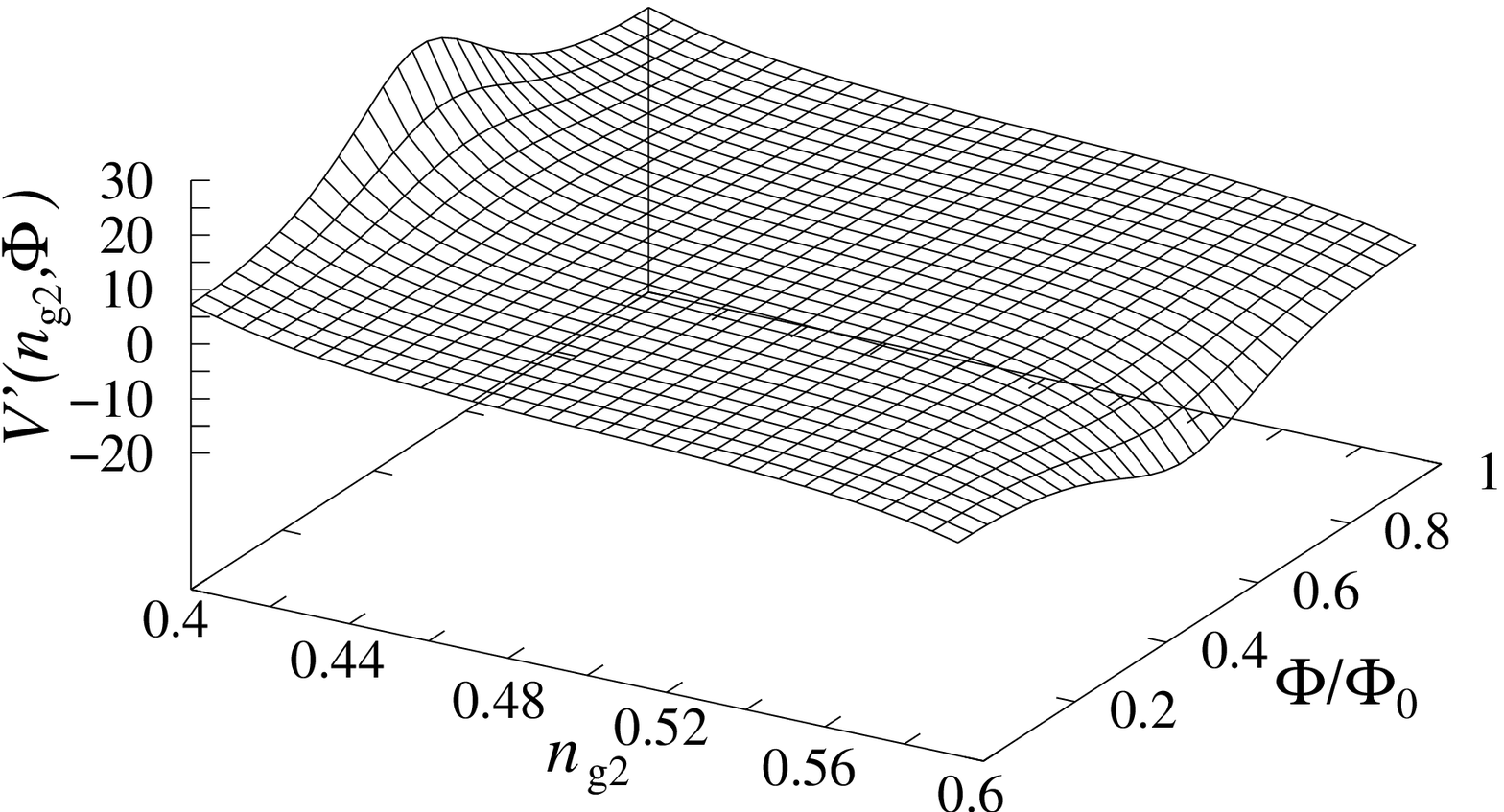}}}}
\caption{Construction of the $\exp (i \phi _2 \sigma _x)$
  rotation. Upper plot -- the degeneracy subspace calculated from the 
  comparison of the energies of the lowest-energy levels; middle
  plot -- the parameter-dependent spectrum of the system (the ground
  state is twofold degenerate); lower plot -- the ``Berry field''. All
  plots taken for the following values of system parameters:
  $E_{c1}/E_m = 0.8$, $E_{c2}/E_m = 1$, $J_{r}^{1(2)}/E_m
  = 0.1$, and $J_{l}^{1(2)}/E_m = 0.05$. Here we assumed for simplicity
  that the dc-SQUIDs 
  are identical so that $J_1 = J_2$ implies $\Phi _1 = \Phi _2 =
  \Phi$. In practice only the first condition needs to be satisfied} 
\label{fi:plotsflip}
\end{figure}

Similarly, we can find the second generic holonomy, namely, $e ^{i
\phi _2 \sigma _x}$. For $J_1 = J_2 \equiv J$ (the coupling should be
complex during the operations, so the fluxes $\Phi _{c1(2)}$ should have
different values than at the starting point, but chosen in this way
that the Josephson energies may be equal),
$E_{01}=E_{10} \equiv E$, the Hamiltonian may be again presented in the
block form, this time in the basis $\{|00 \rangle, |11 \rangle, |\pm
\rangle =(|01\rangle \pm |10 \rangle  )/\sqrt{2}  \}$:
\begin{equation}
  H_2 = \left(\begin{array}{cccc}
      E_{00} & 0 & -{J \over \sqrt{2}}  & 0  \\ 
      0 & E_{11} & -{J \over \sqrt{2}}  & 0 \\
      -{J^* \over \sqrt{2}} & -{J^* \over \sqrt{2}}  & E - J_m/2 & 0  \\
      0 & 0 & 0 & E + J_m/2 
    \end{array}
  \right).
  \label{}
\end{equation}
Comparison of the lowest energy of the upper-left block with $E +
J_m/2$ gives the degeneracy subspace, as shown in the upper plot of
Fig.~\ref{fi:plotsflip}. The middle plot shows the parameter-dependent
spectrum, again the ground state being twofold degenerate. Since the
state $|- \rangle $ does not vary with the parameters, the desired
phase is simply the Berry phase acquired by the ground state in the
upper-left block. The phase here equals
\begin{equation}
  \phi _2 = {1\over 2}
  \int_{S(\Gamma )}
  d n_{g2} d \Phi V'(n_{g2},\Phi).
  \label{}
\end{equation}
The field $V'$ is shown in the lower plot of Fig.~\ref{fi:plotsflip}.

\section{Dissipative effects}\label{sec:diss}

The scheme described above is valid for a system without any strong
constraints on its parameters -- the only idealization so far is the
assumption of perfect symmetry of the small dc-SQUIDs. As already
noted, the operations are performed within a ground state, and the
relaxation does not depopulate the operational subspace. However, as
the degeneracy is maintained by external parameters, we should expect
that the system will suffer from their fluctuations. 
Now we will give qualitative analysis of the noise in the
considered system, and describe how the design can be optimized in
order to suppress the dissipation. In Josephson charge qubits the
Ohmic noise and the low-frequency charge fluctuations
coming from the substrate are usually relatively strong. Since the system is
operated in the 
charge regime, and both sources couple to the charge on the islands
we apply here the spin-boson model with the system Hamiltonian
approximated by its electrostatic part only
\begin{eqnarray}
  \nonumber
  H &=& - {\omega _1 \over 2} \sigma _z^1 - {\omega _1 \over 2}
  \sigma _z^2 + J 
  \sigma _z^1 \sigma _z^2 + \sum_{k}^{} \omega _k a ^{\dagger}_k a_k
  \\ && +
  \left(\sigma _z^1 + \sigma _z^2\right) X
  + \sigma _z^1 Y_1 + \sigma _z^2 Y_2.
  \label{eq:8ra}
\end{eqnarray}  
Here $X, Y_i$ are the bath operators.
We divided the interaction term into the part
describing the common environment for the two islands, and part
with independent environments. 
If now the two islands have equal energy splitting ($\omega _1 =
\omega _2 \equiv \omega _0$), and the dominant interaction with
environment can be modeled by common environment, the subspace
$\{|01 \rangle , |10 \rangle \}$ is well protected (in the limiting
case of completely correlated noise $Y_i \equiv 0$ the states span the
decoherence-free subspace). 
Our goal is then to minimize the uncorrelated noise in
the islands as much as possible. 
To achieve this for the noise coming from the substrate we can, as
discussed in Ref.~\cite{zorin} place the islands close to each other. 
To suppress the  Ohmic noise we could use a common voltage source
controlling the bias of the islands (as in the experiment by
\textcite{pashkin}). As we present here, this may work well also for our
purposes, provided that the islands are nearly identical. We use again
two voltage sources, but now their roles are not symmetric. Instead, the
main,
common source (the fluctuations of which give correlated noise) will be
used to tune the system, while the second, auxiliary, is connected to
one island via very small capacitance (compared to $C_{g1(2)}$) as it
is used to tune a 
small (as explained below) difference between the gate charges
$n_{g1}$ and $n_{g2}$. The resulting noncorrelated noise will be then
much lower than the correlated noise of the main source.

To be more specific, the fluctuations in $n_g$ should be
correlated (rather than in $V_g$) and we require that the
electrostatic properties  
of both islands are nearly identical ($|E_{c1}-E_{c2}|= \delta E_c \ll
E_{c1(2)}$, and $C_{g1}-C_{g2}=\delta C_g \ll C_{g1(2)}$). Then the
common voltage source $V_{g1} = V_{g2}$ gives first of all nearly equal energy
splitting of the islands [$\omega _1 \approx \omega _2$ in
Eq.(\ref{eq:8ra})], and also relates the parameters $n_{g1} \approx
n_{g2}$. We note that for
the starting point as well as during the operation $e^{i \phi _2
  \sigma _x}$ we have $E_{01} = E_{10}$. This for identical islands gives
$n_{g1} = n_{g2}$. Also during the operation $e^{i \phi _1 \sigma _z}$
there is a region in the parameter space in which the relation $n_{g1}
= n_{g1} (n_{g2}, \Phi _2)$ nearly coincides with $n_{g1} = n_{g2}$
[this region is determined by $E_{01}=E_{10} < E_{00},E_{11}$, where
the states $|\bar{0} \rangle $ and $|\bar{1} \rangle $ have equal and
lowest electrostatic energies. The small difference between $n_{g2}$
and $n_{g1}(n_{g2}, \Phi _2)$ is in this region a result of small, but nonzero
Josephson energies].  
We then can use an additional voltage source that is coupled only to
the first island. This voltage will give an additional term
$\delta n_{g1}$ which compensates the effect of small differences in the
properties of the islands, and during the $e^{i \phi _1 \sigma _z}$
operation the small difference $n_{g2} - n_{g1} (n_{g2},
\Phi _2)$.

Since each gate can be performed using various
paths, we may even further improve the fidelity of the operations by
examining the sensitivity of the degeneracy splitting to the
fluctuations of each parameter, and choose paths from the region
least affected by the fluctuations. For instance,
close to the starting point [defined in Eq.(\ref{eq:qra})] the
main voltage $V_g$ fluctuations 
to the first order 
do not shift the system out of the degeneracy subspace. The same
applies to $\Phi _2$ for the $e^{i \phi _1 \sigma _z}$ operation
and to $\Phi $ for $e^{i \phi _2 \sigma _x}$. This can be easily
seen in the upper plots of Figs.~\ref{fi:plotsphase} and
\ref{fi:plotsflip}. Those fluctuations lie within a plane tangential
to the degeneracy subspace. To protect the system also during the
operation we should construct the loops relatively close to the
starting point.


\section{Discussion}\label{sec:concl}

To summarize, we have found that the system properties give us the
opportunity to realize any unitary transformation within a qubit
space using gates of purely geometric origin. Assuming the perfect
performance not affected by the noise we may rely on the simplest
design without strong constraints on the system parameters. However,
such a system composed of two solid-state qubits is
usually strongly probed by the environment which results in fast
dephasing. Our qualitative analysis made for this particular system
specifies the conditions under 
which the system should be well protected even in the presence of noise. 

We have not discussed here the coupling between such elementary blocks, as
the question of applicability of holonomies in quantum computing
remains open and requires detailed analysis of resulting fault tolerance for
individual proposals. However, as the logical basis here is defined by
states differing by the charge configuration of the islands,
such extension based on the method described in Ref.~\cite{faoro}
should be adjustable to our system.

The scheme presented here
may be applied to any system described by the Hamiltonian of two
interacting qubits of the form 
\begin{equation}
  H = B_1 \cdot \sigma ^1 + B_2 \cdot
  \sigma ^2 + J_z \sigma _z^1 \sigma _z^2 + J_{\perp} (\sigma _+^1
  \sigma _-^2 + h.c.),
  \label{eq:6ra}
\end{equation}
provided the parameters are tunable. Since there
are until now many well-developed schemes concerning qubit control as
well as the 
controlled qubit-qubit coupling (usually for the sake of quantum computation),
realization of quantum holonomies may appear very natural and
straightforward in such systems. For the same reason
the model of a four-dimensional quantum system as a candidate to
construct and  observe
holonomies may turn out to be more applicable than the simplest, and
widely used in implementation-independent discussions three-level model
(see for example \cite{pachos}). 

As far as the Josephson-junction systems are considered, we may also use
the device to study the process of the adiabatic charge transport:\cite{pekola} the logical basis 
is spanned by the states with one excess Cooper-pair on one
island and the second with zero. Quite naturally the
second discussed operation, generated by $\sigma_x$, corresponds for
$\phi_2 = \pi /2$ to the charge pumping cycle.\cite{faoro}


\begin{acknowledgments}
The author thanks Yu.~Makhlin for inspiring discussions and numerous
comments on the manuscript, R.~Fazio, J.~Siewert, and G.~Falci for discussions
and useful hints. This work was supported by the
DFG-Schwerpunktprogramm 
``Quanten-Informationsverarbeitung'', EU IST Project SQUBIT, and EC
Research Training Network.
\end{acknowledgments}


\begin{thebibliography}{99}
\bibitem[Berry (1984), M.~V.~Berry]{berry} M.~V.~Berry,
  Proc. R. Soc. London, Ser. A {\bf 392}, 45 (1984).
\bibitem[F.~Wilczek, A.~Zee (1984), F.~Wilczek, A.~Zee]{WZ} F.~Wilczek and
  A.~Zee, Phys. Rev. Lett. {\bf 52}, 2111 (1984). 
\bibitem[Pachos {\it et al.} (1999), J.~Pachos, P.~Zanardi,
  M.~Rasetti]{pachos} J.~Pachos, P.~Zanardi, and M.~Rasetti, Phys. Rev. A {\bf
    61}, 010305(R) (2000); P.~Zanardi and M.~Rasetti, Phys. Lett. A
  {\bf 264}, 94 (1999).
\bibitem[L.~Faoro {\it et al.} (2003), L. Faoro,~J. Siewert,
  R.~Fazio]{faoro} L.~Faoro, J.~Siewert, and R.~Fazio, Phys. Rev. Lett. {\bf
    90}, 028301 (2003).
\bibitem[M.~-S. Choi (2001),M.~-S. Choi]{choi} M.~-S.~Choi, J. Phys.:
  Condens. Matt. 15(46), 7823 (2003).
\bibitem[Pashkin {\it et al.} (2003), Yu.~A.~Pashkin,
  T.~Yamamoto, O.~Astafiev, Y.~Nakamura, D.~V.~Averin,
  J.~S.~Tsai]{pashkin} Yu.~A.~Pashkin,
  T.~Yamamoto, O.~Astafiev, Y.~Nakamura, D.~V.~Averin, and J.~S.~Tsai,
  Nature (London) {\bf 421}, 823 (2003).
\bibitem{duan} L.-M.~Duan and G.-C.~Guo, Phys. Rev. Lett. {\bf 79},
  1953  (1997). 
\bibitem{zanardi} P.~Zanardi and M.~Rasetti, Phys. Rev. Lett. {\bf
    79}, 3306 (1997). 
\bibitem{lidar} D.~A.~Lidar, I.~L.~Chuang, and K.~B.~Whaley, Phys. Rev. Lett. {\bf 81}, 2594 (1998). 
\bibitem{makhlin} Yu.~Makhlin, G.~Sch\"on, and A.~Shnirman, Nature (London)
  {\bf 398}, 305 (1999).
\bibitem{Tinkham} M.~Tinkham, {\it Introduction to Superconductivity},
  2nd ed. (McGraw-Hill, New York, 1996).
\bibitem[Falci {\it et al.} (2000), G.~Falci, R.~Fazio,
  G.~M.~Palma, J.~Siewert, and V.~Vedral]{falci} G.~Falci, R.~Fazio,
  G.~M.~Palma, J.~Siewert, and V.~Vedral, Nature (London) {\bf 407},
  355 (2000). 
\bibitem{zorin} A.~B.~Zorin, F.-J.~Ahlers, J.~Niemeyer, T.~Weimann,
  and H.~Wolf, Phys. Rev. B  {\bf 53}, 13 682 (1996).  
\bibitem{pekola} J. P. Pekola, J. J. Toppari, M. Aunola,
  M. T. Savolainen, and D. V. Averin, Phys. Rev. B {\bf 60}, 9931 (1999). 
\end{thebibliography}
\end{document}